\documentclass{PoS}

\title{VHE gamma-ray observations of transient and variable stellar objects with the MAGIC Telescopes}

\ShortTitle{VHE gamma-ray observations of transient and variable stellar objects with MAGIC}

\author{\speaker{A. Fern\'andez-Barral}\\
        IFAE, Campus UAB, E-08193 Bellaterra, Spain\\
        E-mail: \email{afernandez@ifae.es}}
        
\author{O. Blanch$^{a}$, E. de O\~na Wilhelmi$^{b}$,  D. F. Torres$^{c}$, C. Fruck $^{d}$, D. Hadasch$^{e}$, A. L\'opez-Oramas$^{a,f}$, P. Munar-Adrover$^{g}$ and R. Zanin$^{g}$ for the MAGIC Collaboration\\
       $^a$IFAE, Campus UAB, E-08193 (Bellaterra, Spain)\\ $^b$Institute of Space Sciences, E-08193 (Barcelona, Spain)\\$^c$ICREA and Institute of Space Sciences, E-08193 (Barcelona, Spain)\\$^d$Max-Planck-Institut f\"ur Physik, D-80805 (M\"unchen, Germany)\\$^e$Institute for Cosmic Ray Research, University of Tokyo, 277-8582 (Chiba, Japan)\\ $^f$now at Laboratoire AIM, Service d'Astrophysique, DSM\textbackslash{}IRFU, CEA\textbackslash{}Saclay FR-91191 Gif-sur-Yvette (Cedex, France)\\ $^g$Universitat de Barcelona, ICC, IEEC-UB, E-08028 (Barcelona, Spain)\\}

\abstract{Galactic transients, X-ray and gamma-ray binaries provide a proper environment for particle acceleration. This leads to the production of gamma rays with energies reaching the GeV-TeV regime. MAGIC has carried out deep observations of different transient and variable stellar objects of which we highlight 4 of them here:  LS~I~+61$^{\circ}$303, MWC 656, Cygnus X-1 and SN 2014J. We present the results of those observations, including long-term  monitoring of Cygnus X-1 and  LS~I~+61$^{\circ}$303 (7 and 8 years, respectively). The former is one of the brightest X-ray sources and best studied microquasars across a broad range of wavelengths, whose steady and variable signal was studied by MAGIC within a multiwavelength scenario. The latest results of an unique object, MWC 656, are also shown in this presentation. This source is the first high-mass X-ray binary system detected that is composed of a black hole and a Be star. Finally, we report on the observations of SN 2014J, the nearest Type Ia SN of the last 40 years. Its proximity and early observation gave a remarkable opportunity to study important features of these powerful events.}

\FullConference{The 34th International Cosmic Ray Conference,\\
		30 July- 6 August, 2015\\
		The Hague, The Netherlands}

\begin{document}

\section{Introduction} 
The MAGIC collaboration has developed a large-scale observational program to study gamma-ray binaries and to search for very high energy (VHE, E > 100 GeV) emission from X-ray binaries since the first telescope started taking data in 2004. X-ray binary systems are composed of a compact object, which can be a neutron star (NS) or a black hole (BH), that orbits a stellar companion. These systems can be split into High Mass X-ray Binaries (HMXBs) and Low Mass X-ray Binaries (LMXBs) according to the type of the stellar companion. In this work, we focus on HMXBs. These systems contain a young star of spectral type O or B and a compact object that accretes material from the companion star through an accretion disk or strong stellar wind. X-ray binaries that emit more energy in the gamma-ray range than at X-ray are classified as gamma-ray binaries. In this proceeding we will show the results of 3 variable binary systems: Cygnus X-1, MWC 656 and   LS~I~+61$^{\circ}$303.

Nevertheless, X-ray binaries are not the only variable galactic events expected to emit in the VHE range. Other transient events, like supernovae (SNe), show a proper environment for VHE gamma-ray emission. In this proceeding, we present the results for SN 2014J, the nearest Type Ia SN in the last four decades. Type Ia SNe are extremely luminous stellar explosions whose origin is also a binary system where one of the members, a carbon-oxygen white dwarf, reaches the Chandrasekhar mass limit of 1.4M$_{\odot}$. When this happens, the electron-degenerate core can not support the increase in gravitational pressure, so a thermonuclear explosion of approximately 10$^{51}$ erg takes place. After these powerful events no compact remnant is expected.

All the observations presented in this proceeding were performed with the MAGIC telescopes, a stereoscopic system of two 17 m diameter Imaging Atmospheric Cherenkov Telescopes (IACTs) located in El Roque de los Muchachos on the Canary island of La Palma (28.8$^{\circ}$N, 17.8$^{\circ}$ W, 2225 m a.s.l.). MAGIC was composed of just one stand-alone IACT until 2009 when MAGIC II was constructed. Between summer 2011 and 2012 both telescopes underwent a major upgrade involving the trigger and readout system as well as the MAGIC I camera, enhancing the performance to achieve an integral sensitivity (E>220 GeV) for sources with a Crab Nebula-like spectrum of 0.66$\pm$0.03\% of the Crab Nebula flux in 50 hours of observation in stereoscopic observational mode \cite{p}.  For stand-alone mode (only one telescope operating) the sensitivity above 280 GeV is 1.6\% Crab in 50 hours \cite{Aliu:2008pd}. 

\section{Cygnus X-1}
Cygnus X-1 is one of the most studied and brightest HMXBs \cite{1965Sci...147..394B} of our galaxy, located in the Cygnus region at a distance of 2.15$\pm$0.2 kpc \cite{Ziolkowski:2005ag}. The binary system, composed of a black hole with mass $\sim$15 M$_{\odot}$ \cite{Orosz:2011np} and a 17-31 M$_{\odot}$ O9.7 Iab supergiant companion star \cite{CaballeroNieves:2009mp}, follows a circular orbit of 5.6 days \cite{Brocksopp:1999xs}. It has been firmly established as a microquasar after the detection of a highly collimated (opening angle <2 deg) relativistic (v$\geq$0.6c) one-sided radio-emitting jet \cite{Stirling:2001xb}. The X-ray studies (E< 20keV) revealed that the system behaves as a typical black-hole transient system with the two distinguishable soft and hard state \cite{Fender:2006kw}. 
In 2006, MAGIC observed Cygnus X-1 for 40 hours between June and November with the stand-alone MAGIC telescope (MAGIC I). No significant excess for steady or variable emission was detected, except for one day. During the night of September 24,  which was concurrent with the hard state of the source, an excess of approximately  4.1$\sigma$ after trials was observed, where the significance ($\sigma$) was computed using equation 17 of \cite{Li:1983fv}. The spectrum for that day followed a power-law defined as $dN/(dAdtdE)=(2.3\pm0.6)\cdot10^{-12}(E/1 TeV)^{-3.2\pm0.6}cm^{-2}s^{-1}TeV^{-1}$. After this hint of activity in the gamma-ray regime, MAGIC has carried out observations from 2007 to 2014 focusing on the hard state.
\\
The source was observed for $\sim$80 hours (after data quality cuts) at a zenith range between 6 and 50 deg during 5 observation campaigns in 5 years. The first campaign was performed in a stand-alone mode (with MAGIC I) between June and November 2007 with the tracking mode called wobble-mode \cite{1994APh.....2..137F}. The second one was carried out using only MAGIC I in the on-off tracking mode on July 2008. The campaign from June to October 2009, presents data taken in stand-alone and also stereo mode under wobble-mode. On September 2011, Cygnus X-1 was observed with both telescopes. Finally, on September 2014, the source was observed with same conditions as the previous campaign but during its soft state. 
\\
We performed searches for VHE emission on daily basis (due to the variability of the source), as a function of the different X-ray states and for the full data sample. None of the conditions yielded to significant excess of photons from the source position.

\section{MWC 656}
\begin{figure}[!ht]
 \centering
 \includegraphics[width=0.7\textwidth]{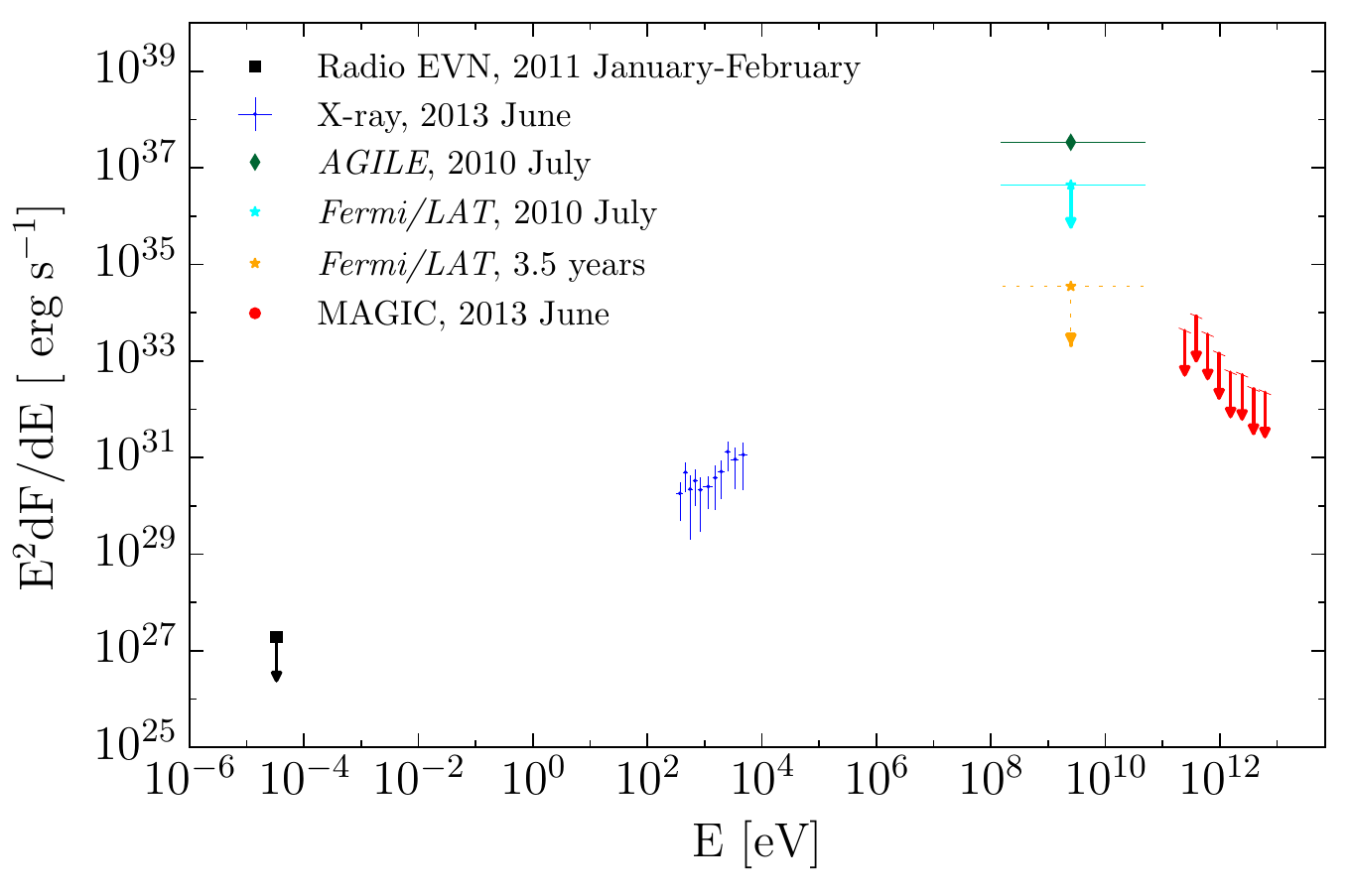}
\caption{Differential ULs from MWC 656 obtained by MAGIC during the 2013 campaign, assuming a photon index of $\Gamma$=2.5. The spectral energy distribution (SED) includes EVN radio ULs \cite{Moldon2012}, the \textit{AGILE} energy flux \cite{2010ATel.2761....1L} and \textit{Fermi}-LAT data taken simultaneously with the AGILE observations \cite{mori13}.}
  \label{fig:diffuls_MWC}
 \end{figure}

MWC 656 is currently the only detected binary composed of a Be star and a BH \cite{Casares:2014gma}, where the authors conclude that this source is a HMXB with a measured BH of 3.8-6.9 M$_{\odot}$. The system is located at a distance of 2.6$\pm$0.6 kpc \cite{Casares:2014gma} and from observed optical photometric modulation, the period of the orbit was determined to be 60.37$\pm$0.04 days \cite{ParedesFortuny:2012ai} with the periastron at $\phi_{per}$=0.01$\pm$0.10 \cite{Casares:2014gma}. On July 2010, AGILE detected a gamma-ray flare locally coincident with the system, which triggered MAGIC observations \cite{2009A&A...502..995T}.
\\
In order to evaluate if the source emits in the VHE gamma-ray regime, MAGIC observed it during two campaigns in a zenith range between 22 and 51 deg. From May to June 2012, 21.3 hours of good quality data in stand-alone mode were taken with MAGIC II between orbital phases $\phi$=0.2 and $\phi$=1.0. On June 2013, the system was observed for $\sim$3.3 hours in stereoscopic mode in the orbital phase range $\phi$=0.0-0.1, just after the periastron. During this last observation epoch, XMM-\textit{Newton} observed the source immediately following MAGIC observations for $\sim$1 hour on June 4th ($\phi$=0.8). No specific information in the X-ray energy range was available for the period of 2012.
\\
The system did not show significant VHE gamma-ray excess in any epoch, either steady or daily basis emission. The integral flux upper limit (UL) for the entire data sample was set at $2.0\cdot10^{-12}cm^{-2}s^{-1}$ at 95\% confidence level (CL) above 300 GeV with a photon index of $\Gamma$=2.5. The data distributed along a phase binning width of 0.1 was also analyzed with no significant emission. The computed differential flux ULs between 245 GeV (energy threshold of the analysis) and 6.3 TeV at 95\% CL, with five bins per decade of energy, are shown in Figure \ref{fig:diffuls_MWC}.  Any potential steady VHE emission is far away from a detectable level with any IACT, although  the possibility of a detection can no be ruled out in the case of a flare occurring at the level of the flux detected by \textit{AGILE}.

\section{LS~I~+61$^{\circ}$303}
 LS~I~+61$^{\circ}$303 is a member of the small group of gamma-ray binaries that has been detected in a very broad wavelength range, from radio up to VHE gamma-rays. The system is composed of a Be star (spectral type B0Ve, \cite{1981PASP...93..486H}) with a circumstellar disk and an unidentified compact object (NS or BH). The orbit that the compact object performs presents an eccentricity of 0.54$\pm$0.03 and a period of 26.4960(28) days obtained by radio analysis \cite{2002ApJ...575..427G}. The periastron phase has been set to $\phi_{per}$=0.23$\pm$0.03. 
The source was detected in HE gamma-ray by \textit{Fermi}-LAT with periodic outbursts around $\phi\sim$0.3-0.45. In the VHE regime, the first detection was performed by MAGIC \cite{magic06}, and confirmed later on by VERITAS \cite{veritas08}, whose periodic peak is detected at $\phi\sim$0.6-0.7 (next to the apastron). Sporadic emission with significant flux at phases $\phi\sim$0.8-1.0 has also been reported by MAGIC ($\sim$4\% Crab flux) \cite{2009ApJ...693..303A} and only detected close to the periastron ($\phi$=0.081) once by VERITAS \cite{2011ApJ...738....3A}.
The non-thermal component of the emission presents periodical outbursts in an orbit-to-orbit base (1667$\pm$8 days, detected in the radio band). This superorbital variability was also found in HE \cite{2013ApJ...773L..35A}.
Despite the multiwavelength information, the nature of its VHE emission is not clearly defined yet. There are three proposed scenarios: the microquasar scenario, due to the observed extended jet-like radio-emitting structure \cite{2004A&A...414L...1M}, the pulsar wind scenario, thanks to the rotating tail-like elongated morphology of overall size 5-10 mas obtained by VLBA \cite{2006smqw.confE..52D} and finally,  LS~I~+61$^{\circ}$303 was proposed to be the first binary system that holds a magnetar \cite{2012ApJ...744..106T}, after the \textit{Swift} detection of two very short-timescale (< 0.1 s) highly-luminous (> 10$^{37}$ ergs$^{-1}$) bursts from the source direction. In this flip-flop magnetar model the pulsar magnetosphere is disrupted close to the periastron due to the circumstellar disk of the Be star (propeller regime) suppressing VHE gamma-ray emission, while in phases next to the apastron the particles can be accelerated up to TeV due to the rotational-powered pulsar (ejector regime). A higher mass-loss rate of the star implies a higher circumstellar disk of the star and vice versa, so that the propeller regime can take place even close to the apastron.  
\\
In order to confirm the superorbital variability in the VHE regime, MAGIC observed LS~I~+61$^{\circ}$303  when the periodical VHE outburst happens, $\phi$=0.55-0.75, from August 2010 to September 2014 in stereoscopic mode (except January 2012, when mono observations with MAGIC II were performed). Archival MAGIC data (from 2006 to 2010, giving rise to a total 8-year campaign) and published VERITAS data have also been used to study this variability. 
The study of the viability of the pulsar wind scenario, and more specifically the flip-flop magnetar model, is carried out searching for (anti-)correlation of the TeV emission and the mass-loss rate of the Be star. To achieve that goal MAGIC performed simultaneous observations with the optical telescope LIVERPOOL (also located in El Roque de los Muchachos) during orbital phases $\phi$=0.8-1.0 where sporadic TeV emission has previously been detected. The mass-loss rate of the star is correlated with the $H_{\alpha}$ line emission, thus the Pearson correlation coefficient and the probabilities of the correlation between the equivalent width (EW), full width half maximum (FWHM) and the velocity of the $H_{\alpha}$ line and the TeV flux were calculated. 
\\
Since  LS~I~+61$^{\circ}$303 presents yearly variability in the peak of the periodical outburst, spectral studies to find evidence for various mechanism of gamma-ray production were carried out. Studies using the entire data set and studies using samples based on superorbital, orbital phase and flux level yield a spectral index compatible with 2.43$\pm$0.04 in general.

\begin{figure}[ht]
 \centering
 \includegraphics[width=0.7\textwidth]{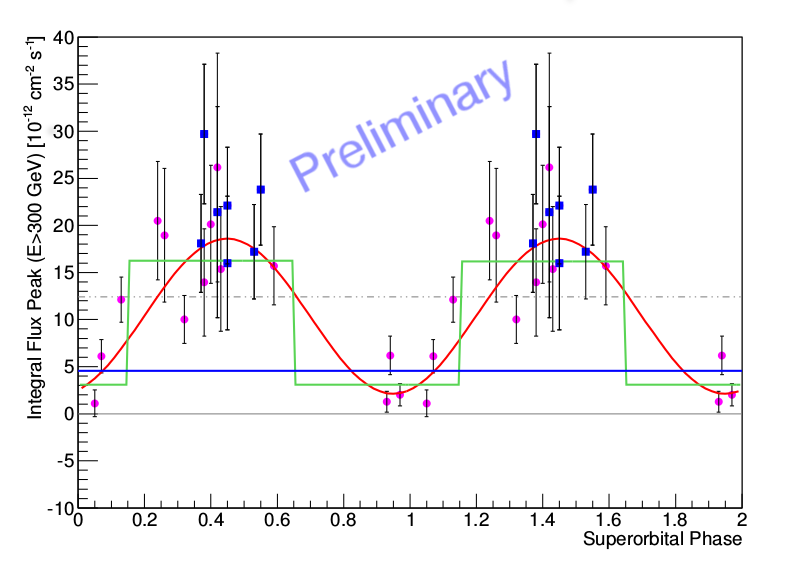}
\caption{Peak flux emitted for each orbital period, for orbital phases 0.5-0.75, in terms of the superorbital phase as defined by radio \cite{2002ApJ...575..427G}. MAGIC (magenta dots) and VERITAS (blue squares) points have been used in this analysis. The fit to a sinusoidal (solid red line), to a step function (solid green line) and with a constant (solid blue line) are also represented. The gray dashed line represents 10\% of the Crab Nebula flux. The gray solid line marks the zero level, just as a reference.}
\vspace{-15pt}
  \label{fig:superorbital_LSI}
 \end{figure}

Figure \ref{fig:superorbital_LSI} shows the data folded in the superorbital period. The probability that the data describes a constant flux is a negligible value of $4.5\cdot10^{-12}$. Assuming a sinusoidal signal, the fit probability reaches 8\%. The data have furthermore been fitted by a step function, resulting in a fit probability of $4.7\cdot10^{-2}$. This shows that the intensity distribution cannot be described only with a high and a low state, but that a gradient is needed. It can be concluded that there is a superorbital signature in the TeV emission of LS~I~+61$^{\circ}$303 and that it is compatible with the $\sim$4.5 year radio modulation seen in other wavelenghts.
No (anti-)correlation has been found between the mass-loss rate of the star and the TeV emission. Nevertheless, the relation between these two parameters can not be either confirmed or denied since the timescale of the observations in optical and TeV differs from minutes to hours, respectively. 

\section{SN 2014J}
\begin{figure}[ht]
		\centering
		\includegraphics[width=0.7\textwidth]{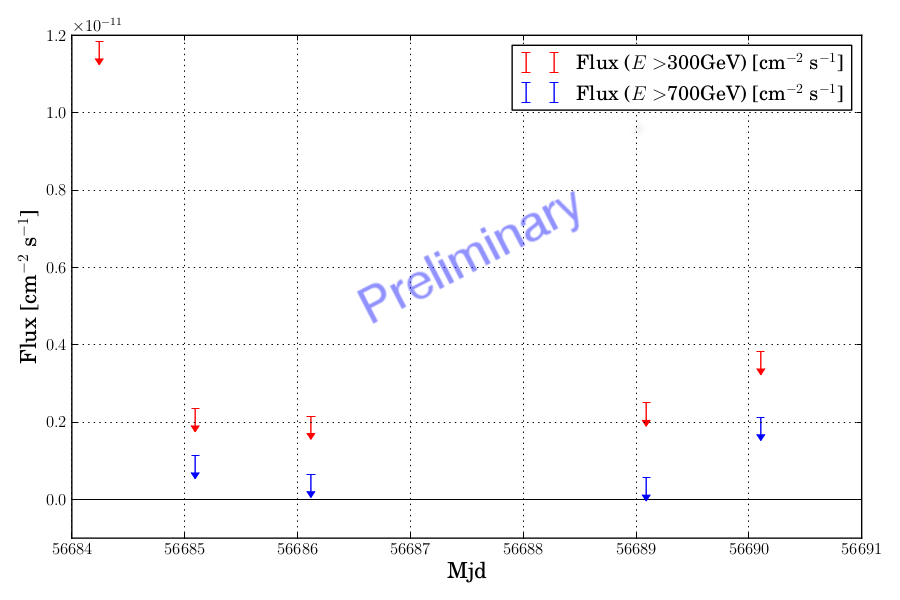} 
		\caption{Daily flux upper limits for gamma-rays from the position of SN 2014J for energies above 300 GeV (red) and 700 GeV (blue). No upper limit is computed for the first day for energies above 700 GeV due to a lack of statistics.}
		\label{fig:SNULs}
\end{figure}

On January 21st, 2014, the UCL Observatory detected SN 2014J at a distance of 3.6 Mpc in the starburst galaxy M82, which was classified by the Dual Imaging Spectrograph on the ARC 3.5m telescope as a Type Ia SN. The proximity of this event gave SN 2014J the title of the nearest SN Type Ia in the last 42 years. Due to its proximity, a multiwavelength follow-up observation campaign was carried out. 
\\
SN 2014J was observed with MAGIC from January 27th to 29th under moderate moonlight conditions and on February 1st and 2nd under dark-night conditions. In total, 6 hours of good-quality data were taken at medium zenith angles, between 40 and 52 deg. 
\\
No signal from the direction of the source was detected. The integral UL for energies above 300 GeV was set to $1.50\cdot10^{-12} cm^{-2}s^{-1}$ with 95\% CL. Taking into account the detection of the host galaxy, M82, reported by VERITAS \cite{2009Natur.462..770V} for energies above 700 GeV, $(3.70\pm0.8_{stat}\pm0.7_{syst})\cdot10^{-13} cm^{-2}s^{-1}$, we have also established an integral UL for this energy range at $3.90\cdot10^{-12} cm^{-2}s^{-1}$ at the same CL. Daily flux ULs for energies above 300 and 700 GeV were also computed and are shown in Figure \ref{fig:SNULs}. 

\section{Conclusions}
MAGIC has performed several observation campaigns on transient and variable stellar objects. Cygnus X-1 was observed from 2007 to 2014 focusing on the hard X-ray spectral state when a detection was expected according to the hint obtained previously by MAGIC. No excess of VHE gamma rays has been found in any period. MWC 656 was observed during the periastron passage, close to the reported AGILE emission, with no significant excess. Integral ULs at energies above 300 GeV have been set for the $\sim$25 hours of good-quality data available for this source.  LS~I~+61$^{\circ}$303 observations revealed that the source displays a superorbital variability consistent with the radio and HE results. In order to probe the flip-flop magnetar model, the (anti-)correlation between the mass-loss rate of the Be companion star and the TeV emission has been studied, although no clear correlation has been found. Observations of SN 2014J have also been presented in this work. No detection in the VHE regime by MAGIC can be reported.

\section{Acknowledgments}
We would like to thank
the Instituto de Astrof\'{\i}sica de Canarias
for the excellent working conditions
at the Observatorio del Roque de los Muchachos in La Palma.
The support of the German BMBF and MPG,
the Italian INFN, 
the Swiss National Fund SNF,
and the ERDF funds under the Spanish MINECO
is gratefully acknowledged.
This work was also supported
by the CPAN CSD2007-00042 and MultiDark CSD2009-00064 projects of the Spanish Consolider-Ingenio 2010 programme,
by grant 268740 of the Academy of Finland,
by the Croatian Science Foundation (HrZZ) Project 09/176 and the University of Rijeka Project 13.12.1.3.02,
by the DFG Collaborative Research Centers SFB823/C4 and SFB876/C3,
and by the Polish MNiSzW grant 745/N-HESS-MAGIC/2010/0.

\bibliographystyle{JHEP}
\bibliography{bibliography_skeleton}

\end{document}